\documentclass[%
 aip,
 amsmath,amssymb,
 reprint,%
 citeautoscript,
]{revtex4-1}

\usepackage{graphicx}%
\usepackage{dcolumn}%
\usepackage{bm}%
\usepackage{braket}
\usepackage{xcolor}

\usepackage[utf8]{inputenc}
\usepackage[T1]{fontenc}
\usepackage{mathptmx}
\usepackage{siunitx}

\newcommand\footnoteref[1]{\protected@xdef\@thefnmark{\ref{#1}}\@footnotemark}
\begin{document}

\title{OrbNet: Deep Learning for Quantum Chemistry Using Symmetry-Adapted Atomic-Orbital Features} %

\author{Zhuoran Qiao} 
\affiliation{%
Division of Chemistry and Chemical Engineering, California Institute of Technology, Pasadena, CA 91125 %
}%

\author{Matthew Welborn} 
\affiliation{%
Entos, Inc., 4470 W Sunset Blvd., Suite 107 PMB 94758, Los Angeles, CA 90027}

\author{Animashree Anandkumar}
\affiliation{%
Division of Engineering and Applied Sciences, California Institute of Technology, Pasadena, CA 91125%
}%

\author{Frederick R. Manby} 
\affiliation{%
Entos, Inc., 4470 W Sunset Blvd., Suite 107 PMB 94758, Los Angeles, CA 90027}

\author{Thomas F. Miller III}
\email{tfm@caltech.edu, tom@entos.ai}
\affiliation{%
Division of Chemistry and Chemical Engineering, California Institute of Technology, Pasadena, CA 91125 %
}%
\affiliation{%
Entos, Inc., 4470 W Sunset Blvd., Suite 107 PMB 94758, Los Angeles, CA 90027}

\date{\today}%
\begin{abstract}
We introduce a machine learning method in which energy solutions 
from the Schrodinger equation are predicted using symmetry adapted atomic orbitals features and a graph neural-network architecture.   \textsc{OrbNet} is shown to outperform existing methods in terms of learning efficiency and transferability for the prediction of density functional theory results while employing low-cost features that are obtained from semi-empirical electronic structure calculations.  For applications to datasets of drug-like molecules, including QM7b-T, QM9, GDB-13-T, DrugBank, and the conformer benchmark dataset of Folmsbee and Hutchison, \textsc{OrbNet} predicts energies within chemical accuracy of DFT at a computational cost that is thousand-fold or more reduced. 
\end{abstract}

\maketitle

\section{Introduction}
The potential energy surface is the central quantity of interest in the modelling of molecules and materials. Calculation of
these energies with sufficient accuracy in chemical, biological, and materials systems is in many -- but not all -- cases adequately described at the level of density functional theory (DFT). 
However, due to its relatively high cost, the applicability of DFT is limited to either relatively small molecules or modest conformational sampling, at least in comparison to force-field and semi-empirical quantum mechanical theories.
A major focus of machine learning (ML) for quantum chemistry has therefore been to  improve the efficiency with which potential energies of molecular and materials systems can be predicted while preserving accuracy.

In the context of quantum chemistry, many applications have focused on the use atom- or geometry-specific feature representations and kernel-based\cite{Bartok2010, rupp2012fast, christensen2020fchl,christensen2019operator,ramakrishnan2015big,Nguyen2018, Fujikake2018,Grisafi2018,zhai2020active} or neural-network (NN) ML architectures.\cite{Smith2017, smiti_transfer_2018, Lubbers, VonLilienfeld2013,hansen2013assessment, Ceriotti2014,Behler2016,kearnes2016molecular,schutt2017quantum,Tuckerman,Smith2017,wu2018moleculenet,Yao2018,Li2018, Zhang2018} 
Recent studies  focus on the  featurization of molecules in abstracted representations -- such as quantum mechanical properties obtained from low-cost electronic structure calculations\cite{mobml1,mobml2,RegressionClustering,neuralxc,deephf} -- and the utilization of novel graph-based neural network\cite{gcn, gat, yang2019analyzing, schutt2017schnet, unke2019physnet, DimeNet, DeepMoleNet} techniques to improve transferability and learning efficiency. 
In this vein, we present a new  approach (\textsc{OrbNet}) based on the featurization of molecules in terms of symmetry-adapted atomic orbitals (SAAOs) and the use of graph neural network methods for deep-learning quantum-mechanical properties. 
We  demonstrate the performance of the new method for the prediction of molecular properties, including the total and relative conformer energies for molecules in  
a range of datasets of organic and drug-like molecules. %
The method enables the prediction of molecular potential energy surfaces with full quantum mechanical accuracy while enabling vast reductions in computational cost; moreover, the method outperforms existing methods in terms of its training efficiency and transferable accuracy across diverse molecular systems.

\section{Method}

\begin{figure*}
    \centering
    \includegraphics[width=0.8\textwidth]{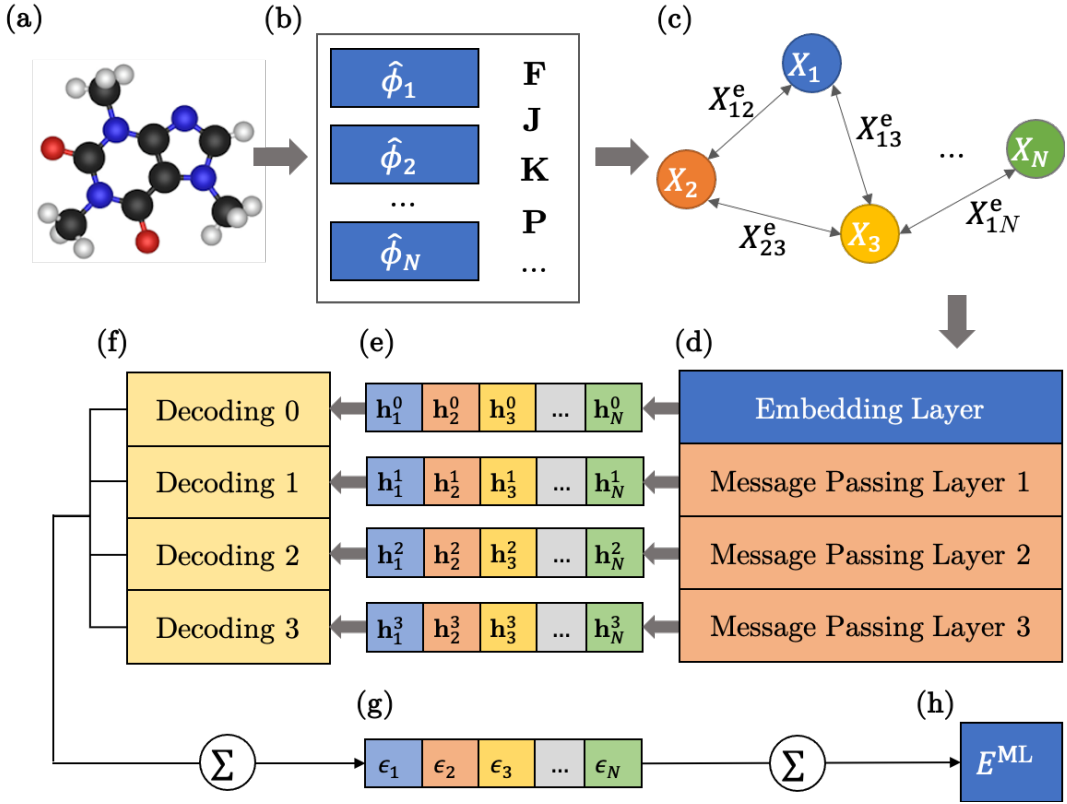}
    \caption{Summary of the \textsc{OrbNet} workflow. (a) A low-cost mean-field electronic structure calculation is performed for the molecular system, and (b) the resulting SAAOs and the associated quantum operators are constructed. (c) An attributed graph representation is  built with node and edge attributes corresponding to the diagonal and off-diagonal elements of the SAAO tensors. (d) The attributed graph is processed by the embedding layer and message passing layers to produce transformed node and edge attributes. (e) The transformed node attributes for the encoding layer and each message passing layer are extracted and (f) passed to MPL-specific decoding networks. (g) The node-resolved energy contributions $\epsilon_{u}$ are obtained by summing the decoding networks outputs node-wise, and (h) the final extensive energy prediction  is obtained from a one-body summation over the nodes. %
    }
    \label{fig:gnn}
\end{figure*}

The target of this  work is to machine-learn a transferable mapping from input feaures values $\{\mathbf{f}\}$ to the regression labels that are quantum mechanical energies,
\begin{equation}
    \label{eq:ML_functional} 
    E \approx %
    E^\mathrm{ML}\left[\{\mathbf{f}\}\right].
\end{equation}
The key elements of \textsc{OrbNet} (Fig.~\ref{fig:gnn}) include the efficient evaluation of the features in the SAAO basis, the utilization of a graph  neural-network architecture with edge and node attention and message passing layers, and a prediction phase that ensures extensivity of the resulting energies. 
We summarize these elements in the current section and discuss the relationship between \textsc{OrbNet} %
and other ML approaches.
Although results in the current paper are presented for the mapping of features from semi-empirical-quality features to DFT-quality labels, the method is  general with respect to the mean-field method used for features (i.e., also allowing for Hartree-Fock, DFT, etc.) and the level of theory used for generating labels (i.e., also allowing for coupled-cluster and other correlated-wavefunction-method reference data). 

\subsection{SAAO Features}
Let $\{\phi^{A}_{n,l,m}\}$ be the set of atomic orbital (AO) basis functions with atom index $A$ and the standard principal and angular momentum quantum numbers,  $n$, $l$, and $m$. Let $\mathbf{C}$ be the corresponding molecular orbital coefficient matrix obtained from a mean-field electronic structure calculation, such as HF theory, DFT, or a semi-empirical method. The one-electron density matrix of the molecular system in the AO basis is then
\begin{equation}
    \label{DM}
    P_{\mu\nu} = 2\sum_{i\in\text{occ}}C_{\mu i}C_{\nu i}
\end{equation}
(for a closed-shell system).
We construct a rotationally invariant symmetry-adapted atomic-orbital (SAAO) basis $\{\hat{\phi}^{A}_{n,l,m}\}$ by diagonalizing diagonal density-matrix blocks associated with indices $A$, $n$, and $l$, such that 
\begin{equation}
    \label{saao}
    \mathbf P^A_{nl}
    \mathbf Y^A_{nl}
    =
    \mathbf Y^A_{nl} \operatorname{diag}(\lambda^A_{nlm})
\end{equation}
where $[\mathbf P^A_{nl}]_{mm'}=P^A_{nlm,nlm'}$.
For s orbitals ($l=0$), this symmetrization procedure is obviously trivial, and can be skipped.
By construction, SAAOs are localized and consistent with respect to geometric perturbations of the molecule, and in contrast with localized molecular orbitals (LMOs) obtained from minimizing a localization %
objective function (Pipek-Mezey, Boys, etc.), SAAOs are obtained by a series of very small diagonalizations, without the need for an iterative procedure. %
The SAAO eigenvectors $\mathbf{Y}^A_{nl}$ are aggregated to form a block-diagonal transformation matrix $\mathbf Y$ that specifies the full transformation from AOs to SAAOs:
\begin{equation} \label{eq:saao_transform}
     |\hat{\phi}_{p} \rangle = \sum_{\mu} Y_{\mu p} |\phi_{\mu} \rangle,
\end{equation}
where $\mu$ and $p$ index the AOs and SAAOs, respectively. %

We employ ML features $\{\mathbf{f}\}$ comprised %
of tensors obtained by evaluating quantum-chemical operators in the SAAO basis. 
Hereafter, all quantum mechanical matrices will be assumed to represented in the SAAO basis, including the density matrix $\mathbf{P}$ and the overlap matrix $\mathbf{S}$.
Following our previous work,\cite{mobml1} the features include expectation values of the %
Fock ($\mathbf{F}$), Coulomb ($\mathbf{J}$), and exchange ($\mathbf{K}$) operators in the SAAO basis. In %
this work, we additionally include the SAAO density matrix, $\mathbf{P}$, the orbital centroid distance matrix, $\mathbf{D}$, the core Hamiltonian matrix, $\mathbf{H}$, and the overlap matrix, $\mathbf{S}$; 
other quantum-mechanical matrix elements are also possible for featurization.

\subsection{Approximated Coulomb and exchange %
SAAO features}

When a semi-empirical quantum chemical theory is employed, %
the computational bottleneck of SAAO feature generation becomes %
the $\mathbf{J}$ and $\mathbf{K}$ terms, due to the need to compute four-index electron-repulsion integrals. We address this problem by introducing a generalized form of the Mataga--Nishimoto--Ohno--Klopman formula, as %
in the sTDA-xTB method,\cite{grimme2013simplified, grimme2016ultra}
\begin{equation}
\label{eq:Ax}
    (pq|rs)^\textrm{MNOK} = \sum_A\sum_B Q_{pq}^{A} Q_{rs}^{B} \gamma_{AB}.
\end{equation}
Here, $A$ and $B$ are atom indices,  $p,q,r,s$ are SAAO indices, and
\begin{equation}
    \gamma_{AB}^{\{\mathbf{J},\mathbf{K}\}}=\left(\frac{1}{R_{AB}^{y_{\{\mathbf{J},\mathbf{K}\}}}+\eta^{-y_{\{\mathbf{J},\mathbf{K}\}}}}\right)^{1/{y_{\{\mathbf{J},\mathbf{K}\}}}},
\end{equation}
where $R_{AB}$ is the distance between atoms $A$ and $B$, $\eta$ is the average chemical hardness for the atoms $A$ and $B$, and $y_{\{\mathbf{J},\mathbf{K}\}}$ are empirical parameters specifying the decay behavior of the damped interaction kernels, $\gamma_{AB}^{\{\mathbf{J},\mathbf{K}\}}$. In this work, we used $y_\mathbf{J}=4$ and $y_\mathbf{K}=10$ similar to which employed in the sTDA-RSH method\cite{risthaus2014excited}. The transition density $Q_{pq}^{A}$ is calculated from a Löwdin population analysis,
\begin{equation}
\label{eq:Lowdin}
    Q_{pq}^{A}=\sum_{\mu \in A} Y'_{\mu p} Y'_{\mu q},
\end{equation}
where the $p$th column of $\mathbf{Y'}=\mathbf{Y}\mathbf{S}^{1/2}$ contains the expansion coefficients for the $p$th SAAO in the symmetrically orthgonalized AO basis. 
This yields  approximated $\mathbf{J}$ and $\mathbf{K}$ matrices for featurization,
\begin{equation}
    J_{pq}^\mathrm{MNOK}=(pp|qq)^\textrm{MNOK} =\sum_{A,B} Q_{pp}^{A} Q_{qq}^{B} \gamma_{AB}^\mathbf{J}
    \label{eq:mnok_j}
\end{equation}
\begin{equation}
    K_{pq}^\mathrm{MNOK}=(pq|pq)^\textrm{MNOK} =\sum_{A, B} Q_{pq}^{A} Q_{pq}^{B} \gamma_{AB}^\mathbf{K}
    \label{eq:mnok_k}
\end{equation}
A naive implementation of Eqs.~\ref{eq:mnok_j}  and \ref{eq:mnok_k} %
is $\mathcal{O}(N^4)$, the leading asymptotic cost. However, this scaling may be reduced to $\mathcal{O}(N^2)$ with negligible loss of accuracy through a tight-binding approximation; for molecules in this study, computation of $\mathbf{J}^\textrm{MNOK}$ and $\mathbf{K}^\textrm{MNOK}$ is not the leading order cost for feature generation and such tight-binding approximation is thus not employed.

\subsection{\textsc{OrbNet}}

\textsc{OrbNet} encodes the molecular system as graph-structured data and utilizes a graph neural network (GNN) machine-learning architecture. %
The GNN represents data as an attributed graph $G(\mathbf{V}, \mathbf{E}, \mathbf{X}, \mathbf{X^e})$, with nodes $\mathbf{V}$, edges $\mathbf{E}$, node attributes $\mathbf{X}: \mathbf{V}\xrightarrow{}\mathbb{R}^{n\times d}$, and edge attributes $\mathbf{X^e}: \mathbf{E}\xrightarrow{}\mathbb{R}^{m\times e}$, where $n=|V|$,  $m=|E|$, and $d$ and $e$ are the number of attributes per node and edge, respectively.

Specifically, \textsc{OrbNet} employs a  graph representation for a molecular system in which node attributes correspond to diagonal SAAO features $X_u=[F_{uu}, J_{uu}, K_{uu}, P_{uu}, H_{uu}]$ and edge attributes  correspond to off-diagonal SAAO features $X^\mathrm{e}_{uv}=[F_{uv}, J_{uv}, K_{uv}, D_{uv}, P_{uv}, S_{uv}, H_{uv}]$. 
By introducing an edge attribute cutoff value for edges to be included, %
non-interacting molecular systems separated at infinite distance are encoded as disconnected graphs, thereby satisfying size-consistency.

The model capacity is enhanced by introducing nonlinear input-feature transformations to the graph representation via radial basis functions,
\begin{equation}
    \label{nrbf}
    \mathbf{h}^{\textrm{RBF}}_{u} = [\phi^\mathrm{h}_1(\tilde{X}_{u}),  \phi^\mathrm{h}_2(\tilde{X}_{u}), ..., \phi^\mathrm{h}_{n_\mathrm{r}}(\tilde{X}_{u})] 
\end{equation}
\begin{equation}
    \label{erbf}
    \mathbf{e}^{\textrm{RBF}}_{uv} = [\phi^\mathrm{e}_1(\tilde{X}^\mathrm{e}_{uv}), \phi^\mathrm{e}_2(\tilde{X}^\mathrm{e}_{uv}),  ...,  \phi^\mathrm{e}_{m_\mathrm{r}}(\tilde{X}^\mathrm{e}_{uv})],
\end{equation}
where $\tilde{\mathbf{X}}$ and $\tilde{\mathbf{X}}^\mathbf{e}$ are $n\times d$ and $m\times e$ matrices with pre-normalized attributes, as described in the Computational Details section.  Sine basis functions $\phi_{n}^\mathrm{h}(r) = \sin(\pi n r)$ are used for node embedding. Motivated by the embedding approach introduced by a recent atom-based GNN study\cite{DimeNet}, we employ 0-th order spherical Bessel functions for edge embedding,
\begin{equation}
    \phi_{m}^\mathrm{e}(r) = j_0^m(r/c_\mathbf{X}) \cdot I_\mathbf{X}(r) = \sqrt{\frac{2}{c_\mathbf{X}}} \frac{\sin(\pi m r / c_\mathbf{X})}{r / c_\mathbf{X}} \cdot I_\mathbf{X}(r),
\end{equation}
where $c_\mathbf{X} \ (\mathbf{X}\in \set{\mathbf{F}, \mathbf{J}, \mathbf{K}, \mathbf{D}, \mathbf{P}, \mathbf{S}, \mathbf{H}})$ is the operator-specific upper cutoff value to $\tilde{X}^\mathrm{e}_{uv}$. To ensure that the feature varies smoothly when a node enters the cutoff, we further introduce the mollifier $I_\mathbf{X}(r)$:
\begin{equation}
    \label{moll}
    I_\mathbf{X}(r) = 
     \begin{cases}
        \exp\left(-\frac{c_\mathbf{X}^2}{(|r|-c_\mathbf{X})^2} + 1\right) &\quad\text{if } 0 \le |r| < c_\mathbf{X} \\
       0 &\quad\text{if } |r| \ge c_\mathbf{X} \\
     \end{cases}
\end{equation}
Note that $\phi_{m}^\mathrm{e}(r)$ decays to zero as an edge approaches the cutoff to ensure size-consistency,
and the mollifier is infinite order differentiable at the boundaries, which eliminates representation noise that can arise from geometric perturbation of the molecule. To enforce that the output is constant at machine precision when adding arbitrary numbers of zero edge features, which is critical for the extraction of analytical gradients and training potential energy surfaces, we also introduced an `auxiliary edge' scheme to be integrated with the message passing mechanism,
\begin{equation}
    \label{aux}
    \mathbf{e}_{uv}^{\textrm{aux}} = \mathbf{W}^{\textrm{aux}} \cdot \mathbf{e}_{uv}^{\textrm{RBF}},
\end{equation}
where $\mathbf{W}^{\textrm{aux}}$ is a trainable parameter matrix. 
The radial basis function embeddings are transformed by %
neural network modules to yield 0-th order node and edge attributes,
\begin{equation}
    \label{h0e0}
    \mathbf{h}_{u}^{0} = \mathrm{Enc}_\mathrm{h}(\mathbf{h}_{u}^{\textrm{RBF}}),\  \mathbf{e}_{uv}^{0} = \mathrm{Enc}_\mathrm{e}(\mathbf{e}_{uv}^{\textrm{RBF}}).
\end{equation}
where $\mathrm{Enc}_\mathrm{h}$ and $\mathrm{Enc}_\mathrm{e}$ are residual blocks\cite{resnet} comprising 3 dense NN layers with skip connection between the first layer's and the last layer's outputs.
In contrast to atom-based message passing neural networks, this additional embedding transformation captures the interactions among the physical operators.

\begin{figure}
    \centering
    \includegraphics[width=\columnwidth]{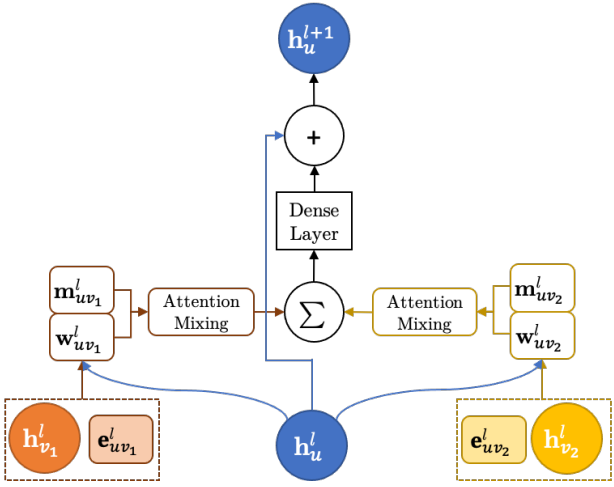}
    \caption{%
    Summary of the \textsc{OrbNet} MPL update.
    For the $l+1$ MPL,  the attributes of a given node (blue) are updated due to interactions with nearest-neighbor  nodes (red and gold), which depend on both the  nearest-neighbor node attributes and the nearest-neighbor edge attributes. 
    The node and edge features (i.e., $\mathbf{h}_{u}^{l}$, $\mathbf{h}_{v}^{l}$, and $\mathbf{e}_{uv}^{l})$ combine to produce a message $\mathbf{m}^{l}_{uv}$ (Eq. \ref{eq:msg}) and multi-head attention score $\mathbf{w}^{l}_{uv}$ (Eq. \ref{eq:attn}) which undergo attention mixing.  The attention-weighted messages from each nearest-neighbor node and edge are combined and passed into a dense layer, the result of which is added to the original node attributes to perform the update (Eq. \ref{eq:mpl}). 
    }
    \label{fig:mpl}
\end{figure}

The node and edge attributes are updated via the Transformer-motivated\cite{transformer} message passing mechanism in Fig.~\ref{fig:mpl}.
For a given %
message passing layer (MPL) %
$l+1$, the information carried by each edge is
encoded into a message function $\mathbf{m}_{uv}^{l}$ and associated attention weight $\mathbf{m}_{uv}^{l}$, and is accumulated into node features through a graph convolution operation. The overall message passing mechanism is given by:
\begin{equation}
    \mathbf{h}_u^{l+1} = \mathbf{h}_{u}^l + \sigma\Big(\mathbf{W}_\mathrm{h}^{l} \cdot \big[\bigoplus\limits_{i} \big(\sum_{v\in N(u)} w^{l,i}_{uv} \cdot \mathbf{m}_{uv}^l\big)\big] + \mathbf{b}_\mathrm{h}^{l}\Big),
    \label{eq:mpl}
\end{equation}
where $\mathbf{m}_{uv}^{l}$ is the message function on each edge, computed from the Hadamard product between linearly-projected node attributes and the edge attribute 
\begin{equation}
    \mathbf{m}_{uv}^{l} = \sigma(\mathbf{W}_\mathrm{m}^{l,2} \cdot \big[(\mathbf{W}_\mathrm{m}^{l,1} \cdot \mathbf{h}_u^l) \odot (\mathbf{W}_\mathrm{m}^{l,1} \cdot \mathbf{h}_v^l) \odot \mathbf{e}_{uv}^l \big] + \mathbf{b}_\mathrm{m}^{l})
    \label{eq:msg}
\end{equation}
and the convolution kernel weights, $w^{l,i}_{uv}$, are evaluated as (multi-head) %
attention scores\cite{gat} to characterize the relative importance of an orbital pair,
\begin{equation}
    w^{l,i}_{uv}= \sigma_\mathrm{a}( \sum [(\mathbf{W}_\mathrm{a}^{l,i} \cdot \mathbf{h}_u^l) \odot (\mathbf{W}_\mathrm{a}^{l,i} \cdot \mathbf{h}_v^l) \odot \mathbf{e}_{uv}^l  \odot \mathbf{e}^{\textrm{aux}}_{uv}]/n_{\mathrm{e}}),
    \label{eq:attn}
\end{equation}
where the summation is applied over the elements of the vector in the summand. 
Here, the index $i$ specifies a single attention head, and $n_{\mathrm{e}}$ is the dimension of hidden edge features $\mathbf{e}^{l}_{uv}$, 
$\bigoplus$ denotes a vector concatenation operation, $\odot$ denotes the Hadamard product, and $\cdot$ denotes the matrix-vector product.

The edge attributes are updated according to 
\begin{equation}
    \mathbf{e}_{uv}^{l+1} = \sigma(\mathbf{W}_\mathrm{e}^{l} \cdot \mathbf{m}_{uv}^{l} + \mathbf{b}_\mathrm{e}^{l}),
\end{equation}
$\mathbf{W}_\mathrm{m}^{l,1}$, $\mathbf{W}_\mathrm{m}^{l,2}$, $\mathbf{W}_\mathrm{h}^{l}$, $\mathbf{W}_\mathrm{e}^{l}$, $\mathbf{b}_\mathrm{m}^{l}$, $\mathbf{b}_\mathrm{h}^{l}$, $\mathbf{b}_\mathrm{e}^{l}$ and $\mathbf{a}^{l}$ are %
MPL-specific trainable parameter matrices, $\mathbf{W}_\mathrm{a}^{l,i}$ are %
MPL- and attention-head-specific trainable parameter matrices, $\sigma(\cdot)$ is an activation function with a normalization layer, and $\sigma_\mathrm{a}(\cdot)$ is the activation function used for generating attention scores.

The decoding phase of \textsc{OrbNet} (Fig.~\ref{fig:gnn}f-h) is designed to ensure the size-extensivity of energy predictions. %
The employed mechanism outputs node-resolved energy contributions for %
the embedding layer ($l=0$) and all MPLs ($l=1, 2, ..., L$) to predict the energy components associated with all nodes and MPLs. 
The final energy prediction $E^{\textrm{ML}}$ is obtained by first summing over $l$ (Fig. \ref{fig:gnn}g) for each node $u$ and then
performing a one-body sum over nodes (i.e., orbitals) (Fig. \ref{fig:gnn}h), such that  
\begin{equation} 
    E^\mathrm{ML} = \sum_{u \in \mathbf{V}} \epsilon_u = \sum_{u \in \mathbf{V}} \sum_{l=0}^L \mathrm{Dec}^l(\mathbf{h}_u^l),
\end{equation}
where the decoding networks $\mathrm{Dec}^l$ are multilayer perceptrons.

\subsection{\textsc{Comparison with other methods that use quantum mechanical features}}

Several ML methods have been developed for the prediction of high-level (i.e., coupled-cluster) correlation energies based on quantum mechanical features from a  mean-field-level  (i.e., HF theory or DFT) electronic structure calculation.\cite{mobml1, mcgibbon2017improving, deephf, margraf2018making} 
An example from our own work includes the  molecular-orbital-based machine-learning (MOB-ML) approach to predict molecular properties using localized molecular orbitals for input feature generation.\cite{mobml1, mobml2, RegressionClustering}
Localized molecular orbitals are obtained via an orbital localization procedure (Boys, IBO, etc), with the orbitals obtained from a mean-field electronic structure calculation. Feature vectors are then calculated for diagonal and off-diagonal molecular orbital pairs from matrix elements of the molecular orbitals with respect to various operators (i.e., Fock, Coulomb, and exchange operators) within the  basis and using a feature sorting scheme. Gaussian-process or clustering-based regressors are trained for the pair correlation energy labels associated to the MOB feature vectors. 

Closer in spirit to \textsc{OrbNet} are  NeuralXC\cite{neuralxc} and DeePHF\cite{deephf}, which %
employ AO-based features obtained from electronic structure calculations to perform the regression and prediction of molecular energies. Both NeuralXC and DeePHF utilize the electronic density and orbitals obtained from either a Hartree-Fock (HF) (in DeePHF) or low-level density functional theory (DFT) (in NeuralXC) calculation using cc-pVDZ or larger atomic-orbital basis sets. %
 However, these methods typically require a mean-field calculation in the same-sized atomic orbital basis set as that of the high-level correlation method (i.e., they do not directly make predictions on the basis of features that are obtained in a minimal basis), and they have not been applied for the prediction of DFT-quality results on the basis of lower-level semi-empirical methods, such as GFN-xTB, as is done here.

In terms of featurization methods,  \textsc{OrbNet} differs from NeuralXC and DeePHF by providing a more information-rich quantum mechanical representation.
Unlike NeuralXC, \textsc{OrbNet} avoids shell-averaging of the AOs, and  unlike both NeuralXC and DeePHF, \textsc{OrbNet} includes all  off-diagonal operator matrix elements (including both intra- and inter-atom elements, as well as intra- and inter-shell elements) within the features, thereby preserving  information content while also enabling description of long-range  contributions. 
Unlike DeePHF, \textsc{OrbNet}   includes interactions between different shells on the same atom and avoids the need for a pre-determined weighting function based on inter-atomic distances.
\textsc{OrbNet} additionally includes quantum-chemical matrices including $\mathbf{F}, \mathbf{J}, \mathbf{K}$ %
which are valuable components for energy prediction tasks. 
Other differences arise in the way in which rotational invariance is enforced within the features.
In NeuralXC, the rotational invariance of the features is guaranteed by summing all sub-shell components of the AO-projected density $d^{nl} = \sum_{m=-l}^{l} c_{nlm}^2$ (i.e. the trace of the local density matrix), such that the information content is not preserved.
In DeePHF, the rotational invariance of the features is enforced by using the eigenvalues of the local density matrix instead of the trace to build the feature vector for each shell.
By contrast, \textsc{OrbNet} achieves the rotational invariance of features through the use of SAAOs, which involve no loss of information content.

In terms of  ML regression methods,  \textsc{OrbNet} also differs from NeuralXC and DeePHF.
For NeuralXC, the ML regression is performed using a Behler-Parrinello\cite{Behler2007} type dense neural network. 
Similarly, for DeePHF, the ML regression is performed using a dense neural network, with the labels associated with a one-body summation over the atoms to yield the total correlation energy. 
In contrast, \textsc{OrbNet} uses a GNN for the ML regression.  Specifically, we report results using a multi-head graph attention mechanism and residual blocks to improve the representation capacity of the model, to learn complex chemical environments. Unlike the pre-tuned aggregation coefficients in DeePHF, \textsc{OrbNet} also offers a flexible framework for learning orbital interactions and could be naturally transferred to downstream tasks.

\section{Computational details}
\label{sec:comp}
Results are presented for the QM7b-T dataset\cite{mobml2,mobdatasets} 
(which has seven conformations for each of 7211 molecules\cite{VonLilienfeld2013} with up to seven heavy atoms of type C, O, N, S, and Cl),
the QM9 dataset\cite{qm9} (which has locally optimized geometries for 133885 molecules with up to nine heavy atoms of type C, O, N, and F),
the GDB-13-T dataset\cite{mobml2,mobdatasets} 
(which has six conformations for each of 1000 molecules from the GDB-13 dataset\cite{GDB-13} with up to thirteen heavy atoms of type C, O, N, S, and Cl), 
DrugBank-T 
(which has six conformations for each of 168 molecules from the DrugBank database\cite{law2014drugbank} with between fourteen and 30 heavy atoms of type C, O, N, S, and Cl),
and the 
Hutchison conformer dataset from Ref.~\citenum{Hutch} 
(which has up to 10 conformations for each of 622 molecules with between nine and 50 heavy atoms of type C, O, N, F, P, S, Cl, Br, and I). 
Except for DrugBank-T, all of these datasets have been described previously; thermalized geometries from the DrugBank dataset are sampled at 50 fs intervals from \textit{ab initio} molecular dynamics trajectories performed using the B3LYP\cite{Vosko1980,Lee1988,Becke1993,Stephens1994}/6-31g*\cite{Hariharan1973} level of theory and a Langevin thermostat\cite{Bussi2007} at 350 K.
The structures for the DrugBank-T dataset are provided in the Supporting Information, and all other employed datasets are already available online.\cite{mobdatasets,qm9,Hutch}
For results reported in Section \ref{sec:qm9}, the pre-computed DFT labels from Ref.~\citenum{qm9} were employed.
For results reported in Section \ref{sec:conf}, all DFT labels were computed using the $\omega$B97X-D functional\cite{wb97xd} with a Def2-TZVP AO basis set\cite{def2tzvp} and using density fitting\cite{Polly2004} for both the 
Coulomb and exchange integrals using the Def2-Universal-JKFIT basis set;\cite{def2-universal-jkfit} %
these calculations are performed using \textsc{Psi4}.\cite{Psi4}  %
Semi-empirical calculations are performed using the GFN1-xTB method\cite{gfn1} using the \textsc{Entos Qcore}\cite{entos} package, which is also employed for the SAAOs feature generation. 
For the results presented in this work, we train \textsc{OrbNet} models using the following  training-test splits of the datasets.
For results on the QM9 dataset, we removed 3054 molecules due to a failed a geometric consistency check, as recommended in Ref.~\citenum{qm9};
we then randomly sampled 110000 molecules for training and used %
10831 molecules for testing. 
The training sets of 25000 and 50000 molecules in section \ref{sec:qm9} are subsampled from the 110000-molecule dataset.
For the QM7b-T dataset, two sets of training-test splits are generated; for the model trained on the QM7b-T dataset only (Model 1 in Section \ref{sec:conf}), we randomly selected 6500 different molecules (with 7 geometries for each) from the total 7211 molecules for training, holding out 500 molecules (with 7 geometries for each) for testing; for Models 2-4 in Section \ref{sec:conf}, we used a 361-molecule subset of this 500-molecules set for testing, and we used the remaining 6850 molecules of QM7b-T for training.  
For the  GDB13-T dataset, we randomly sampled 948 different molecules (with 6 geometries for each) for training, holding out 48 molecules (with 6 geometries for each) for testing. 
For the DrugBank-T dataset, we randomly sampled 158 different molecules (with 6 geometries for each) for training, holding out 10 molecules (with 6 geometries for each) for testing.
No training on the Hutchison conformer dataset was performed, as it was only used for transferability testing.
Since none of the training datasets for \textsc{OrbNet} included molecules with elements of type P, Br, and I, we  excluded the molecules in the Hutchison dataset that included elements of these types for the reported tests (as was also done in Ref.~\citenum{Hutch} and in Fig.~\ref{fig:hutch}  for the ANI methods).
Moreover, following Ref.~\citenum{Hutch}, we excluded sixteen molecules due to missing DLPNO-LCCSD(T) reference data;  an additional eight molecules were excluded on the basis of DFT convergence issues for at least one conformer using \textsc{Psi4}. 
The specific molecules that appear in all training-test splits %
are listed in the Supporting Information.

Table \ref{table:hp} summarizes the hyperparameters used for training  \textsc{OrbNet} for the reported results. 
We perform a pre-transformation on the input features from $\mathbf{F}$, $\mathbf{J}$, $\mathbf{K}$, $\mathbf{D}$, $\mathbf{P}$, $\mathbf{H}$ and $\mathbf{S}$ to obtain $\tilde{\mathbf{X}}$ and $\tilde{\mathbf{X}}^\mathrm{e}$: We normalize all diagonal SAAO tensor values $X_{uu}$ to range $[0, 1)$ for each operator type to obtain $\tilde{X}_{u}$; for off-diagonal SAAO tensor values, we take $\tilde{X}_{uv}=-\ln(|X_{uv}|)$ for $\mathbf{X} \in \Set{\mathbf{F}, \mathbf{J}, \mathbf{K}, \mathbf{P}, \mathbf{S}, \mathbf{H}}$, and $\tilde{D}_{uv}=D_{uv}$.
The model hyperparameters are selected within a limited search space; the cutoff hyperparameters $c_\mathbf{X}$ are obtained by examining the overlap between feature element distributions between the QM7b-T and GDB13-T datasets. The same set of hyperparameters is used throughout this work, thereby providing a universal model.

\begin{table*}[]
\setlength{\extrarowheight}{0.05cm}
\begin{tabular}{ccc}
\hline\hline
Hyperparameter    & Meaning   & Value or name  \\ \hline
$n_\mathrm{r}$   & Number of basis functions for node embedding & 8 \\ \hline
$m_\mathrm{r}$   & Number of basis functions for edge embedding & 8 \\ \hline
$n_\mathrm{h}$   & Dimension of hidden node attributes & 256 \\ \hline
$n_\mathrm{e}$   & Dimension of hidden edge attributes & 64  \\ \hline
$n_\mathrm{a}$   & Number of attention heads          & 4   \\ \hline
$L$              & Number of message passing layers   & 3   \\ \hline
$L_\mathrm{enc}$ & Number of dense layers in $\mathrm{Enc}_\mathrm{h}$ and $\mathrm{Enc}_\mathrm{e}$ & 3   \\ \hline
$L_\mathrm{dec}$ & Number of dense layers in a decoding network  & 4   \\ \hline
                 & Hidden dimensions of a decoding network & 128, 64, 32, 16 \\ \hline
$\sigma$         & Activation function & Swish   \\ \hline
$\sigma_\mathrm{a}$ & Activation function for attention generation  & TanhShrink          \\ \hline
$\gamma$           & Batch normalization momentum  & 0.4 \\ \hline
$c_\mathbf{F}$     & Cutoff value for $\tilde{F}_{uv}$   & 8.0  \\ \hline
$c_\mathbf{J}$     & Cutoff value for  $\tilde{J}_{uv}$ & 1.6 \\ \hline
$c_\mathbf{K}$     & Cutoff value for  $\tilde{K}_{uv}$ & 20.0 \\ \hline
$c_\mathbf{D}$     & Cutoff value for $\tilde{D}_{uv}$ & 9.45  \\ \hline
$c_\mathbf{P}$     & Cutoff value for $\tilde{P}_{uv}$ & 14.0 \\ \hline
$c_\mathbf{S}$     & Cutoff value for $\tilde{S}_{uv}$ & 8.0  \\ \hline
$c_\mathbf{H}$     & Cutoff value for $\tilde{H}_{uv}$ & 8.0  \\ \hline\hline
\end{tabular}
\caption{Model hyperparameters employed in \textsc{OrbNet}. All cutoff values are in atomic units.}
\label{table:hp}
\end{table*}

To provide additional regularization for predicting energy variations from the configurational degree of freedom, we performed training on loss function of the form %
\begin{eqnarray}
    \mathcal{L}(\hat{\mathbf{E}}, \mathbf{E}) &=& (1-\alpha) \sum_{i}\mathcal{L}_\mathrm{2}(\hat{E}_i, E_i)\nonumber\\
    &+& \alpha \sum_{i} \mathcal{L}_\mathrm{2}(\hat{E}_i-\hat{E}_{t(i)}, E_i-E_{t(i)}).
    \label{eq:confloss}
\end{eqnarray}
For a conformer $i$ in a minibatch, we randomly sample another conformer $t(i)$ of the same molecule to be paired with $i$ to evaluate the relative conformer loss $\mathcal{L}_\mathrm{2}(\hat{E}_i-\hat{E}_{t(i)}, E_i-E_{t(i)})$, putting additional penalty on the prediction errors for configurational energy variations. $\mathbf{E}$ denotes the ground truth energy values of the minibatch, $\hat{\mathbf{E}}$ denotes the model prediction values of the minibatch, and $\mathcal{L}_\mathrm{2}$ denotes the L2 loss function $\mathcal{L}_\mathrm{2}(\hat{y},y)=||\hat{y}-y||_{2}^{2}$. For all models in %
Section~\ref{sec:qm9}, we choose $\alpha=0$ as only the optimized geometries are available; for models in Section~\ref{sec:conf}, we choose $\alpha=0.9$ for all training setups.

All models are trained on a single Nvidia Tesla V100-SXM2-32GB GPU using %
the Adam optimizer.\cite{kingma2014adam} For all training runs, we set the minibatch size to 64 and use a cyclical learning rate schedule\cite{smith2019super} that performs a linear learning rate increase from $\num{3E-5}$ to $\num{3E-3}$ for the initial 100 epochs, a linear decay from $\num{3E-3}$ to $\num{3E-5}$ for the next 100 epochs, and an exponential decay with a factor of $0.9$ every epoch for the final 100 epochs. Batch normalization\cite{ioffe2015batch} %
is employed before every activation function $\sigma$ except for that used in the attention heads, $\sigma_\mathrm{a}$.

\section{Results}
\label{sec:results}

We present results that focus on the prediction of accurate DFT energies using input features obtained from the GFN1-xTB method.\cite{gfn1}  The GFN family of methods\cite{gfn1,gfn0,gfn2} have proven to be extremely useful for the simulation of large molecular system (1000s of atoms or more) with time-to-solution for energies and forces on the order of seconds.  However, this applicability can be limited by the accuracy of the semi-empirical method,\cite{jiang2020nuclear,Hutch} thus creating a natural opportunity for ``delta-learning'' the difference between the GFN1 and DFT energies on the basis of the GFN1 features. 
Specifically, we consider regression labels associated with the   difference between high-level DFT and the GFN1-xTB total atomization energies, 
\begin{equation}
E^{\textrm{ML}} \approx E^{\textrm{DFT}} - E^{\textrm{GFN1}} - \Delta E^{\textrm{fit}}_\textrm{atoms},
\end{equation}
where the last term is the sum of differences for the isolated-atom energies between DFT and GFN1 as determined by a linear model.
This approach yields the direct ML prediction of total DFT energies, given the results of a GFN1-xTB calculation.

\subsection{The QM9 dataset}
\label{sec:qm9}

\begin{table*}
\begin{ruledtabular}
\setlength{\extrarowheight}{0.1cm}
\begin{tabular}{ccccccccc}
Training size & SchNet\cite{schutt2017schnet} & PhysNet\cite{unke2019physnet} & PhysNet-ens5\cite{unke2019physnet} & DimeNet\cite{DimeNet} &DeepMoleNet\cite{DeepMoleNet} & \textsc{OrbNet} & \textsc{OrbNet}-ens5\\ \hline
25,000 & - & - & -& -& -& \textbf{11.6} & \textbf{10.4} \\ \hline
50,000 & 15 & 13 & 10 & - & - & \textbf{8.22} & \textbf{6.80}\\ \hline
110,000 & 14 & 8.2 & 6.1 & 8.02 & 6.1 & \textbf{5.01} & \textbf{3.92}\\
\end{tabular}
\end{ruledtabular}
\caption{MAEs  (reported in meV) for predicting the QM9 dataset of %
total energies at the B3LYP/6-31G(2df,p) level of theory.
Results from the current work are reported for a single model (\textsc{OrbNet}) and with ensembling over 5 models (\textsc{OrbNet}-ens5). 
}
\label{table:qm9}
\end{table*}

We begin with a broad comparison of recently introduced ML methods for the %
total energy task, $U_0$, from the widely studied QM9 dataset.\cite{qm9} 
QM9 is composed of organic molecules with up to nine heavy atoms at locally optimized geometries, so this test (Table \ref{table:qm9}) examines the expressive power of the ML models for systems in similar chemical environments. 
Results for \textsc{OrbNet} are presented both without ensemble averaging of independently trained models (i.e., predicting only on the basis of the first of trained model) and with ensemble averaging the results of five independently trained models (\textsc{OrbNet}-ens5).  As observed previously,\cite{unke2019physnet} ensembling helps in this and other learning tasks, reducing the \textsc{OrbNet} prediction error by approximately 10-20\%. %

Also included in the table are previously published methods utilizing graph representations of atom-based features, including SchNet\cite{schutt2017schnet}, PhysNet\cite{unke2019physnet}, DimeNet\cite{DimeNet}, and DeepMoleNet\cite{DeepMoleNet}. 
We note that DimeNet employs a directional message passing mechanism and PhysNet and DeepMoleNet employ supervision based on prior physical information to improve the model transferability, which could also be employed within \textsc{OrbNet}; 
it is clear that without these additional strategies  and even without model ensembling, \textsc{OrbNet} provides greater accuracy and learning efficiency than all previous deep-learning methods.

\subsection{Transferability and Conformer Energy Predictions}
\label{sec:conf}

A more realistic and demanding test of ML methods is to train them on datasets of relatively small molecules (for which high-accuracy data is more readily available) and then to test on datasets of larger and more diverse molecules.  This provides useful insight into the transferability of the ML methods and the general applicability of the trained models.

\begin{figure}
    \centering
    \includegraphics[width=\columnwidth]{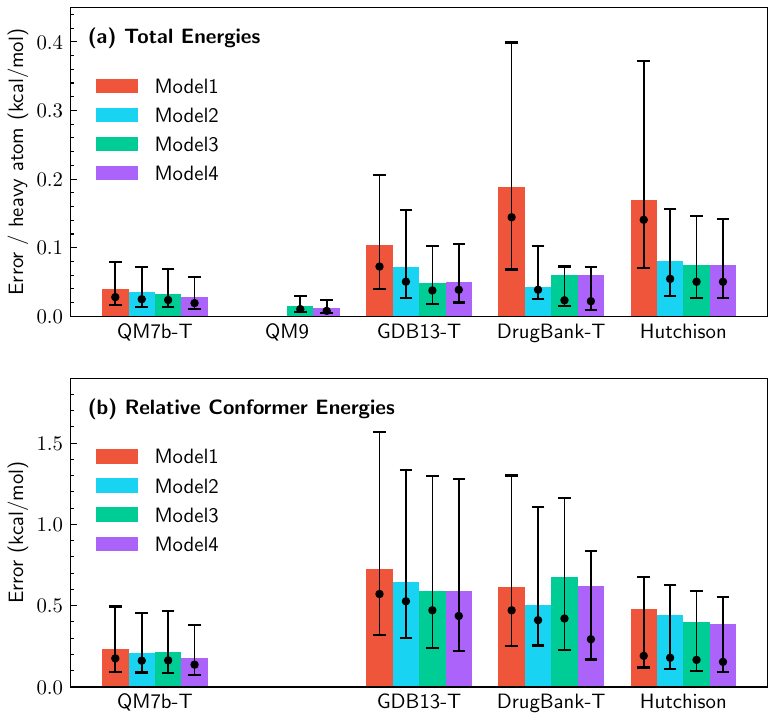}
    \caption{Prediction errors for (a) molecule total energies and (b) relative conformer energies performed using \textsc{OrbNet} models trained using various datasets. 
    The mean absolute error (MAE) is indicated by the bar height, the median of the absolute error is indicated by a black dot, and the 
    the first and third quantiles for the absolute error are indicated as the lower and upper bars. 
   Model 1 uses training data from QM7b-T; Model 2 additionally includes training data from GDB13-T and DrugBank-T; Model 3 additionally includes training data from QM9; and Model 4 additionally includes ensemble averaging over five independent training runs. Testing is performed on data that is held-out from training in all cases. 
    Training and prediction employs energies at the $\omega$B97X-D/Def2-TZVP level of theory.
    All energies in kcal/mol.
    }
    \label{fig:confmaes}
\end{figure}

To this end, we investigate the performance of \textsc{OrbNet} on a series of dataset containing organic and drug-like molecules.
Fig.~\ref{fig:confmaes} presents results in which \textsc{OrbNet} models are trained with increasing amounts of data.  Using the training-test splits described in Section \ref{sec:comp}, Model 1 is trained using data from only the QM7b-T dataset; Model 2 is trained using data from the QM7b-T, GDB13-T, and DrugBank-T datasets; Model 3 is trained using data from the QM7b-T, QM9, GDB13-T, and DrugBank-T datasets; and Model 4 is obtained by ensembling five independent training runs with the same data as used for Model 3.  Predictions are made for total energies (Fig.~\ref{fig:confmaes}A) and relative conformer energies (Fig.~\ref{fig:confmaes}B) for held-out molecules from each of these datasets, as well as for the Hutchison conformer dataset.

As expected, it is seen from Fig.~\ref{fig:confmaes} that the \textsc{OrbNet} predictions improve with additional data and with ensemble modeling. 
The median and mean of the absolute errors consistently decrease from Model 1 to Model 4 except for a non-monotonicity in the DrugBank-T MAE, likely due to the relatively small size of that dataset.
It is nonetheless striking that Model 1, which includes only data from QM7b-T yields relative conformer energy predictions on the DrugBank-T and Hutchison datasets (which include molecules with up to 50 heavy atoms) with an accuracy that is comparable to the more heavily trained models. 
Note that all of the \textsc{OrbNet} models predict relative conformer energies with MAE and median prediction errors that are well within the 1 kcal/mol threshold of chemical accuracy, across all four test datasets. 
Predictions for QM9 using Models 1 and 2 are not included, since QM9 includes F atoms whereas the training data in those models do not; relative conformer energies are not predicted for QM9 since they are not available in this dataset.
Although total energy prediction error for the \textsc{OrbNet} is slightly larger per heavy atom on the Hutchison dataset than for the other datasets, the relative conformer energy prediction error for the Hutchison dataset is slightly smaller than for  GDB13-T and DrugBank-T; this is due to the fact that the Hutchison dataset involves locally minimized conformers that are less spread in energy per heavy atom than the conformers of the thermalized datasets.  This relatively small energy spread among conformers in the Hutchison dataset is a realistic and challenging aspect of drug-molecule conformer-ranking prediction, which we next consider.

\begin{figure}
    \centering
    \includegraphics[width=\columnwidth,trim={0 6cm 0 6cm}]{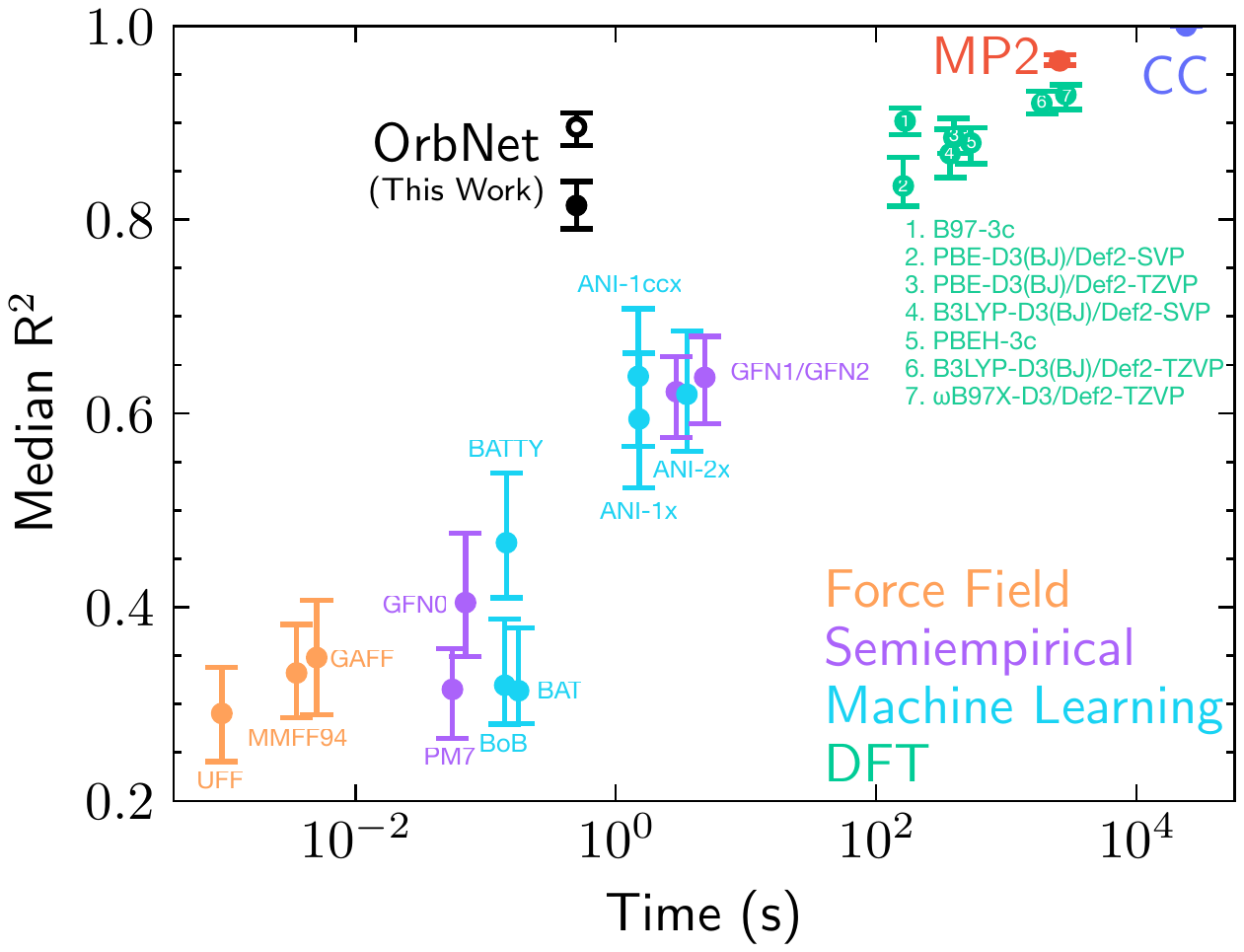}
    \caption{Comparison of the accuracy/computational-cost tradeoff for a range of potential energy methods for the Hutchison conformer benchmark dataset.
    Aside from the \textsc{OrbNet} results (black), all data was previously reported in Ref.~\citenum{Hutch}, with median R$^2$ values for the predicted conformer energies computed %
    relative to  DLPNO-CCSD(T) reference data and with computation time evaluated on a single CPU core.
    The \textsc{OrbNet} results (black) are obtained using Model 4 (i.e., with training data from QM7b-T, GDB13-T, DrugBank-T, and QM9 and with ensemble averaging over five independent training runs).
    The solid black circle plots the median R$^2$ value from the \textsc{OrbNet} predictions relative to  DLPNO-CCSD(T) reference data, as for the other methods.
    The open black circle  plots the median R$^2$ value from the \textsc{OrbNet} predictions relative to the $\omega$B97X-D/Def2-TZVP reference data against which the \textsc{OrbNet} model was trained. Error bars correspond to the 95\% confidence interval, determined by statistical bootstrapping.}
    \label{fig:hutch}
\end{figure}

Figure \ref{fig:hutch} presents a direct comparison of the accuracy and computational cost of \textsc{OrbNet} in comparison to %
a variety of 
other force-field, semiempirical, machine-learning, DFT, and wavefunction methods, as compiled in Ref.~\citenum{Hutch}.
For the Hutchison conformer dataset of drug-like molecules which range in size from nine to 50 heavy atoms, the accuracy of the various methods was evaluated  using the median R$^2$ of the predicted conformer energies in comparison to DLPNO-CCSD(T) reference data and with computation time evaluated on a single CPU core.\cite{Hutch}

The \textsc{OrbNet} conformer energy predictions (Fig.~\ref{fig:hutch}, black) are reported using Model 4 (i.e., with training data from QM7b-T, GDB13-T, DrugBank-T, and QM9 and with ensemble averaging over five independent training runs).
The solid black circle indicates the  median R$^2$ value (0.81) of the \textsc{OrbNet} predictions relative to the DLPNO-CCSD(T) reference data, as for the other methods; this point provides the most direct  comparison to the accuracy of the other methods.
The open black circle indicates the  median R$^2$ value (0.90) of the \textsc{OrbNet} predictions relative to the  $\omega$B97X-D/Def2-TZVP reference data against which the model was trained; this point indicates the accuracy that would be expected of the Model 4 implementation of \textsc{OrbNet} if it had employed coupled-cluster training data rather than DFT training data.  
We performed timings for \textsc{OrbNet} on a single core of an Intel Core i5-1038NG7 CPU @ 2.00GHz, finding that the \textsc{OrbNet} computational cost is dominated by the GFN1-xTB calculation for the feature generation.  
In contrast to Ref.~\citenum{Hutch} which used the \textsc{xtb} code of Grimme and coworkers\cite{xtbcode}, we used \textsc{Entos Qcore} for the GFN1-xTB calculation calculations.  We find the reported timings for GFN1-xTB to be surprisingly slow in  Ref.~\citenum{Hutch}, particularly in comparison to the GFN0-xTB timings.
For  GFN0-xTB, our  timings with \textsc{Entos Qcore} are very similar to those reported in Ref.~\citenum{Hutch}, which is sensible given that the method involves no self-consistent field (SCF) iteration.
However, whereas Ref.~\citenum{Hutch} indicates GFN1-xTB timings that are 43-fold slower than GFN0-xTB, we find this ratio to be only 4.5 with  \textsc{Entos Qcore}, perhaps due to differences of SCF convergence.  To account for the issue of code efficiency in the GFN1-xTB implementation and to control for the details of the single CPU core used in the timings for this work versus in Ref.~\citenum{Hutch}, we normalize the \textsc{OrbNet} timing reported in Fig.~\ref{fig:hutch} with respect to the GFN0-xTB timing from Ref.~\citenum{Hutch}. The CPU neural-network inference costs for \textsc{OrbNet} are negligible contribution to this timing.

The results in Fig.~\ref{fig:hutch} make clear that \textsc{OrbNet} enables the prediction of relative conformer energies  for drug-like molecules with an accuracy that is comparable to DFT but with a computational cost that is 1000-fold reduced from DFT to realm of semiempirical methods.  Alternatively viewed, the results  indicate that \textsc{OrbNet} provides dramatic improvements in prediction accuracy over currently available ML and semiempirical methods for realistic applications, without significant increases in computational cost.

\section{Conclusions}

Electronic structure methods typically face a punishing trade-off between the %
prediction accuracy of the method and its computational cost, across all areas of the chemical, biological, and materials sciences.
We present a new machine-learning method with the potential to substantially shift that trade-off in favor of \emph{ab initio}-quality accuracy at low %
computational cost.
\textsc{OrbNet} utilizes a graph neural network architecture to predict high-quality electronic-structure energies on the basis of features obtained from low-cost/minimal-basis mean-field electronic structure methods.
The method is demonstrated for the case of predicting $\omega$B97X-D/Def2-TZVP energies using GFN1-xTB input features, although it is completely general with respect to both the choice of high-level (including correlated wavefunction) method  used for generating reference data and the choice of mean-field method used for feature generation. 
In comparison to state-of-the-art GNN methods for the prediction of total molecule energies for the QM9 dataset, it is shown that \textsc{OrbNet} provides a 33\% improvement in prediction accuracy with the same amount of data relative to the next-most accurate method (DeepMoleNet).\cite{DeepMoleNet}
And in comparison to the wide array of methods used for predicting relative conformer energies in a realistic and diverse dataset of drug-like molecules, as compiled by Folmsbee and Hutchison,\cite{Hutch} it is shown that \textsc{OrbNet} provides a prediction accuracy that is similar to DFT and much improved over existing ML methods, but at a computational cost that is reduced by at least three orders of magnitude relative to DFT. 
Natural future directions for development will include the expansion of \textsc{OrbNet} to a broader set of chemical elements, incorporation of  directional message-passing and 
model supervision using  prior physical information,\cite{DimeNet,DeepMoleNet,unke2019physnet}
and end-to-end refitting of the semi-empirical method used for feature generation.\cite{Li2018,zhou2020gpu}

\section{Supplemental Material}

The supplemental material includes the structures for the DrugBank-T dataset, as well as specification of molecules that appear in all training-test splits for the trained models.

\begin{acknowledgments}
The authors thanks Lixue Sherry Cheng for providing geometries for the DrugBank-T dataset and Anders Christensen for helpful comments on the manuscript.
Z.Q. acknowledges graduate research funding from Caltech. 
T.F.M. and A.A. acknowledge partial support from the Caltech DeLogi fund, and A.A. acknowledges support from a Caltech Bren professorship. 
\end{acknowledgments}

\section*{Data Availability Statement}
The data that supports the findings of this study are available within the article and its supplementary material.

\bibliography{main}%

\end{document}